\documentclass[structabstract]{aa}  
\usepackage{graphicx}
\usepackage{txfonts}
\begin{document}
   \title{The abundance of C$^{18}$O and HDO in the envelope and hot core of the intermediate mass protostar NGC~7129~FIRS~2 \thanks{{\it Herschel} is an ESA space observatory with science instruments provided by European-led Principal Investigator consortia and with important participation from NASA.}}

   \author{A. Fuente\inst{1}
          \and
          P. Caselli\inst{2}
        \and
         C. M$^{\textrm{c}}$Coey\inst{3}
        \and 
        J. Cernicharo\inst{4} 
       \and
        D. Johnstone\inst{5,6}
        \and
        M. Fich\inst{3}
       \and
	T. van Kempen\inst{7,8}
       \and
        E. van Dishoeck\inst{8,9}
       \and
        U. Y{\i}ld{\i}z\inst{8}
       \and
        R. Visser\inst{8,10}
       \and
       L. Kristensen\inst{8}
	\and
	T. Alonso-Albi
          \inst{1}
          \and
	F. Herpin
         \inst{10,11}
         \and
         S. Tisi\inst{3}
   }

   \institute{Observatorio Astron\'omico Nacional (OAN,IGN), Apdo 112, E-28803 Alcal\'a de Henares (Spain)
              \email{a.fuente@oan.es}
             \and
             School of Physics \& Astronomy, E.C. Stoner Building, The University of Leeds, Leeds LS2 9JT, UK 
            \and
            Department of Physics and Astronomy, University of Waterloo, Waterloo, Ontario, N2L 3G1, Canada
                      \and
           Centro de Astrobiolog\'{\i}a (CSIC/INTA), Laboratory of Molecular Astrophysics, Ctra. Ajalvir km. 4, E-28850, Torrej\'on de Ardoz, Spain
            \and            
     	     Department of Physics \& Astronomy, University of Victoria, Victoria, BC, V8P 1A1, Canada
	    \and
            National Research Council of Canada, Herzberg Institute of Astrophysics, 5071 West Saanich Road,
            Victoria, BC, V9E 2E7, Canada
           \and
           Joint ALMA Offices, Alonso de C\'ordova 3107, Vitacura, Casilla 19001, Santiago 19, Chile
	\and
        Leiden Observatory, Leiden University, P.O. Box 9513, 2300 RA Leiden, Netherlands
        \and
        Max-Planck-Institut f\"ur extraterrestrische Physik, Garching, Germany
        \and
        Department of Astronomy, University of Michigan, 500 Church Street, Ann Arbor, MI 48109-1042, USA
       \and
       Universit\'e de Bordeaux, Observatoire Aquitain des Sciences de l'Univers, 33271, Floirac Cedex, France
        \and
       CNRS, UMR 5804, Laboratoire d'Astrophysique de Bordeaux, 2 rue de l'Observatoire, BP 89, 33271, Floirac Cedex, France
}

   \date{Received September 15, 1996; accepted March 16, 1997}

 
  \abstract
   {NGC~7129~FIRS~2 is a young intermediate-mass (IM) protostar,  which is associated with two energetic bipolar outflows and displays clear signs of the presence of a hot core. It has been extensively observed with ground based telescopes and within the WISH Guaranteed Time Herschel Key Program.}
   {This paper is dedicated to the modeling of the C$^{18}$O and HDO lines in NGC~7129~FIRS~2. Our goal is to investigate the chemistry in the envelope and 
hot core of this IM protostar.}
   {We present new observations of the C$^{18}$O 3$\rightarrow$2 and the HDO  3$_{12}$$\rightarrow$2$_{21}$ lines towards NGC~7129~FIRS~2. Combining these observations with Herschel data and modeling their emissions, we constrain the C$^{18}$O and HDO abundance profiles across the protostellar envelope. In particular, we derive the
abundance of C$^{18}$O and HDO in the hot core.}
   {The intensities of the C$^{18}$O lines are well reproduced assuming that the C$^{18}$O abundance decreases through the protostellar envelope from the outer edge towards the centre until the point where the gas and dust reach the CO evaporation temperature ($\approx$20-25 K) where the C$^{18}$O is released back to the gas phase. Once the C$^{18}$O is released to the gas phase, 
the modelled C$^{18}$O abundance is found to be $\approx$1.6$\times$10$^{-8}$, which is a factor of 10 lower than the reference abundance. This result is supported by the non-detection of C$^{18}$O 9$\rightarrow$8, which proves that even in the hot core (T$_k$$>$100~K) the CO abundance must be 10 times lower than the reference value. Several scenarios are discussed to explain this C$^{18}$O deficiency.  One possible explanation is that during the pre-stellar and protostellar phase, the CO is removed from the grain mantles by reactions to form more complex molecules. Our HDO modeling shows that the emission of HDO  3$_{12}$$\rightarrow$2$_{21}$ line is maser and comes from the hot core (T$_k$$>$100~K).
Assuming the physical structure derived by Crimier et al. (2010),
we determine a HDO abundance of $\sim$0.4$-$1$\times$10$^{-7}$ in the hot core of this IM protostar.}
   {Herschel data combined with ground based observations have allowed us to estimate the C$^{18}$O and HDO abundance in the protostellar envelope and hot 
core of an IM protostar. The HDO abundance in the hot core is $\sim$0.4$-$1$\times$10$^{-7}$,
similar to that found in the hot corinos NGC~1333~IRAS~2A and IRAS~16293$-$2422.
The C$^{18}$O abundance, at $\approx$1.6$\times$10$^{-8}$, is a factor of 10 lower than the reference value. 
}


   \maketitle

\section{Introduction}

CO is the most abundant molecule in the gas phase after molecular hydrogen, and the main
coolant in molecular clouds (Goldsmith \& Langer 1978).  In order for star formation to occur, efficient cooling is essential during the 
early collapse stages.  As a consequence, knowledge of the CO abundance is key toward understanding the evolution of star forming regions (Visser et al. 2011).

Despite decades of extensive observation, there are important limitations in our knowledge 
of the CO chemistry. In particular the role of ice
chemistry remains unclear (Takahashi \& Williams 2000, Mu{\~n}oz-Caro et al. 2010, Garrod \& Pauly 2011,{\"O}berg et al. 2011, Fayolle et al. 2011). Water ice is formed very easily in molecular clouds, as
it requires low gas densities and only moderately low temperatures (Hollenbach et al. 2009).
A thin layer of water ice is formed on dust grains in cold molecular clouds.
In contrast, CO is exclusively formed in gas phase. During pre-stellar collapse, the increasing density
causes CO to accrete onto icy grains, creating a new layer dominated mainly by CO.
Once the protostellar core is formed, the dust is heated up to several hundreds of K and 
the ice is evaporated releasing the molecules back to the gas phase (Viti et al. 2004). Between collapse and core formation, surface reactions 
on the dust grains may change the initial composition of the ice giving rise to new species (Garrod et al. 2008). 
Although there is mounting evidence of
the key role of the surface grain chemistry in the gas chemical composition of hot cores and corinos (Bottinelli et al. 2004ab, 2007; Bisschop et al. 2007; {\"O}berg et al. 2010), our knowledge of the
ice chemistry and the interplay between ice and gas phases in these warm regions suffers large uncertainties.

Intermediate-mass young stellar objects (IMs) are the precursors of stars in the 2$-$8~M$_\odot$ mass range (Herbig Ae/Be stars). 
They share some characteristics with massive young stellar objects
(clustering, PDRs) but many are located in less complex regions and closer to the Sun (d$<$1~kpc), which 
allows determination of the physical and chemical structure of their envelopes at similar spatial scales as
can be done for low mass protostars. IMs are interesting because they constitute the link between the low-mass and high-mass range, in the sense that
they cover an intermediate range of luminosities, densities and temperatures. These different physical conditions are expected to produce a 
different chemistry. In addition, study of IMs is essential 
to establish the initial conditions of the planet-formation process since Herbig Ae stars are the precursor of Vega-type debris disks.

Continuum observations, coupled with molecular line spectra, have recently allowed us to study the physical and chemical structure 
of a sample of 5 IM protostars (Crimier et al. 2010; Alonso-Albi et al. 2010). Crimier et al. (2010) derived the temperature-density profiles 
based on the SED and the SCUBA maps at 450~$\mu$m and 850~$\mu$m. In the case of NGC~7129~FIRS~2, they also took into account the interferometric observations of 
the continuum emission at mm wavelengths published by Fuente et al. (2005a). Adopting these profiles, Alonso-Albi et al. (2010) modelled the chemistry of these envelopes
to fit the radial distribution of the integrated intensity emission of the C$^{18}$O 1$\rightarrow$0, N$_2$H$^+$ 1$\rightarrow$0 and 
N$_2$D$^+$ 2$\rightarrow$1 ( 3$\rightarrow$2 in the case of FIRS~2) lines observed with the IRAM 30m telescope. Surprisingly, the standard model failed to reproduce the integrated intensity maps of all the IMs.  In order to fit to the observations of NGC ~7129~FIRS~2, Cepheus E-mm and CB3, the C$^{18}$O abundance wrt. H$_2$ (in this paper all molecular abundances are wrt. H$_2$)  in the CO evaporation region (where T$_k$$>$25~K) had to be decreased by a factor of 10 relative to the reference value of 1.6$\times$10$^{-7}$ (derived using the expression of Wilson \& Matteucci (1992) assuming a galactrocentric distance of 8.5~Kpc and [$^{16}$O]/[$^{18}$O]=540). These three sources had the highest deuterium fractionation, with [N$_2$D$^+$]/ [N$_2$H$^+$]$>$0.01,  and are thus presumably the youngest of the sample. 
In the case of  Serpens~FIRS~1 and IC~1396~N, the observations were fitted by decreasing the C$^{18}$O abundance by the same factor
but only in the hot core region, i.e. for gas and dust temperatures $>$100~K. In these sources, [N$_2$D$^+$]/ [N$_2$H$^+$]$<$0.01 and the low C$^{18}$O abundance was interpreted as 
the consequence of photodissociation by the UV photons emitted by the central star. This work was based on low excitation lines whose emission is
dominated by the external part of the envelope. The observation of the high-J C$^{18}$O lines was required to confirm our results.

Water is an important species that has a large influence on the chemical and physical evolution of protostellar objects. Unfortunately, its main isotope 
is difficult to observe from the ground because of the Earth's atmosphere (Cernicharo et al. 1990, 1994; Gonz\'alez-Alfonso et al. 1995). 
A good alternative is offered by its 
isotopologue HDO, for which several lines are observable from the ground. One such line,  HDO 3$_{12}$$\rightarrow$2$_{21}$ line at 225 GHz, is typically very bright in protostars because of a maser effect  taking place in the dense and warm inner regions of the protostellar envelope (Ceccarelli et al. 2010).
This line is therefore a good tracer of hot 
cores/corinos and disks (Parise et al. 2005, Liu et al. 2011). 
The detection of HDO in addition of H$_2$O allows estimation of the HDO/H$_2$O ratio. Although the processes leading to the water deuteration are not fully understood, 
it is clearly related to grain surface chemistry and reveals the chemical and physical history of the protostar (Cazaux et al. 2011).

Herschel (Pilbratt et al. 2010) provides the first opportunity to observe high excitation lines of CO (and isotopes)
towards young stellar objects (YSOs). Y{\i}ld{\i}z et al. (2010) modelled the high-J CO (up to J=10$\rightarrow$9), $^{13}$CO and C$^{18}$O lines in the low mass protostar
NGC~1333~IRAS~2A and found they required a CO
abundance in the evaporation zone to be a factor of 3$-$5 lower than the reference value. In this paper we model the high-J C$^{18}$O lines in the IM protostar
NGC~7129~FIRS~2. 
In addition, we present new HDO 3$_{12}$$\rightarrow$2$_{21}$ observations with the 30m telescope.
The abundances of these molecules provide important hints towards understand the chemistry of the hot cores that are the precursors of
proto-planetary disks.

\subsection{A prototypical young intermediate-mass (IM) protostar: NGC~7129~FIRS~2}

With a luminosity of 430 L$_\odot$, NGC~7129~FIRS~2 (hereafter FIRS~2) lies near the middle of the IM luminosity range.
FIRS~2 is the prototypical young IM protostar with clear signs of hot core chemistry (Fuente et al. 2005a), energetic outflows (Fuente et al. 2001) and relatively large deuterium 
fractions (Fuente et al. 2005b; Alonso-Albi et al. 2010). It is located at a distance of 1250$\pm$50~pc from the Sun and belongs to the group of 
IMs in which Alonso-Albi et al. (2010) found a low C$^{18}$O abundance in the CO evaporation region.

The launch of the Herschel Space Observatory (Pilbratt et al. 2010) opened a new observational window in the far-IR, where water lines and high-J CO rotational transitions are present.  
FIRS~2 was observed within the Water In Star-forming regions with Herschel (WISH) key program (van Dishoeck et al. 2011). Johnstone et al. (2010a) used HIFI (de Graauw et al. 2010) data to 
put constraints on the water vapour abundance in the FIRS~2 envelope ($\sim$10$^{-7}$) and observed broad H$_2$O lines likely produced along the outflow. 
Fich et al. (2010) used PACS (Poglitsch et al. 2010) spectroscopy to measure highest excitation lines of CO (from J=14$\rightarrow$13 to J=33$\rightarrow$32) and 
H$_2$O (transitions from E$_u$ = 114 K to 1071 K) and characterized the shocked molecular gas.
This paper aims at determining the C$^{18}$O abundance profile in the envelope of FIRS 2, which can be constrained by combining 
ground-based observations of C$^{18}$O J=1$\rightarrow$0 (Alonso-Albi et al. 2010) and J=3$\rightarrow$2 (this paper) with the Herschel/HIFI observations of 
C$^{18}$O J=5$\rightarrow$4 (Johnstone et al. 2010a) and J=9$\rightarrow$8  (this paper). The large range of excitation and comparable angular resolutions allow us to better explore basic 
chemical processes which determine the CO  abundance across the whole envelope. Ground based observations of the HDO 3$_{12}$$\rightarrow$2$_{21}$ spectrum are also presented. This line is used to derive the HDO abundance in the hot core which complements previous water study based on HIFI observations (Johnstone et al. 2010a).


\onlfig{1}{
\begin{figure*}
\includegraphics{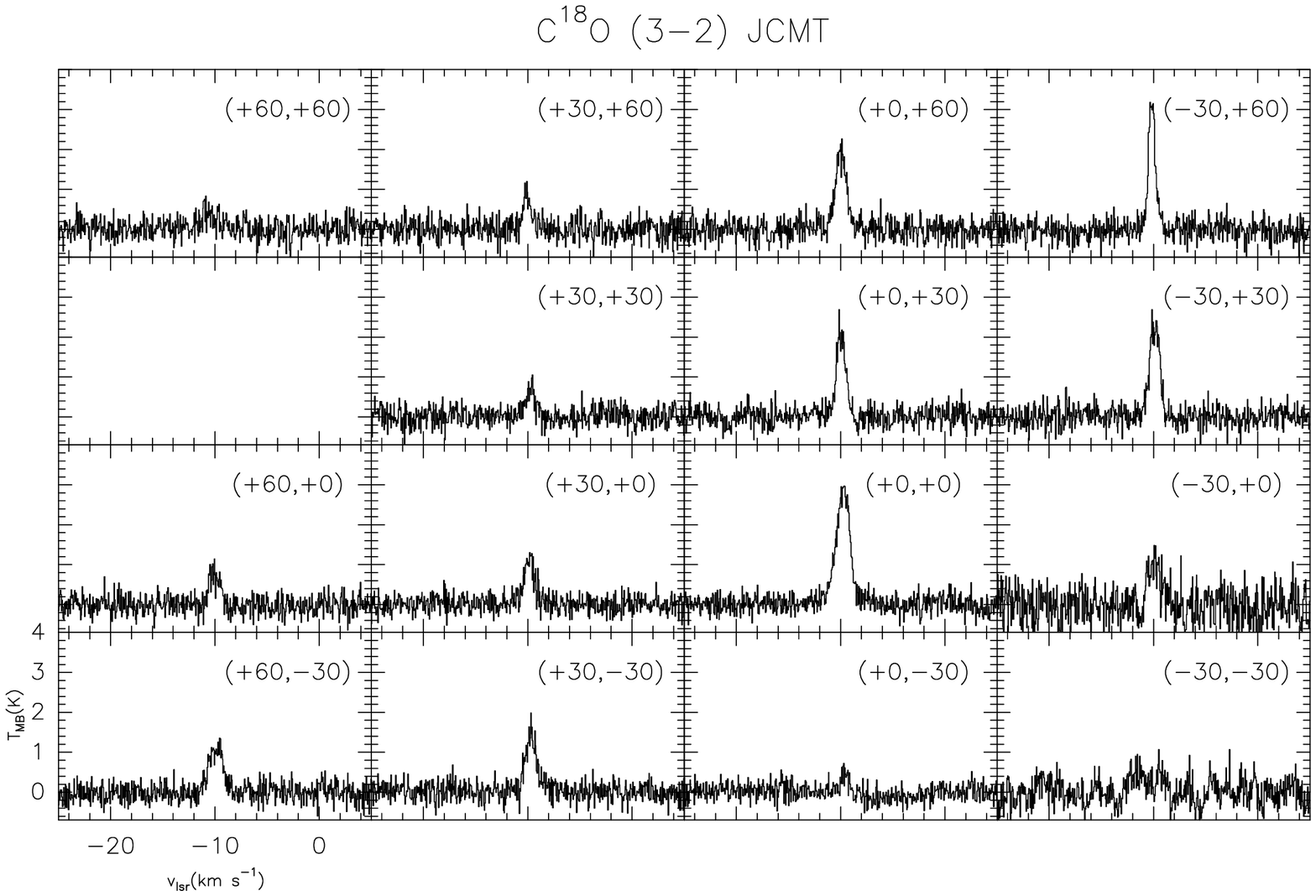}
\vspace{10.0cm}
      \caption{Spectral map of the C$^{18}$O 3$\rightarrow$2 line observed with the JCMT telescope. The intensity scale is main brightness temperature. The (0,0) position corresponds 
to 21$^{\rm h}$ 43$^{\rm m}$ 01\fs7 63\degr 03\arcmin 23\arcsec (J2000). One spectrum was lost because of instrumental problems.}
         \label{Fig 1}
 \end{figure*}
}
\begin{figure*}
\includegraphics{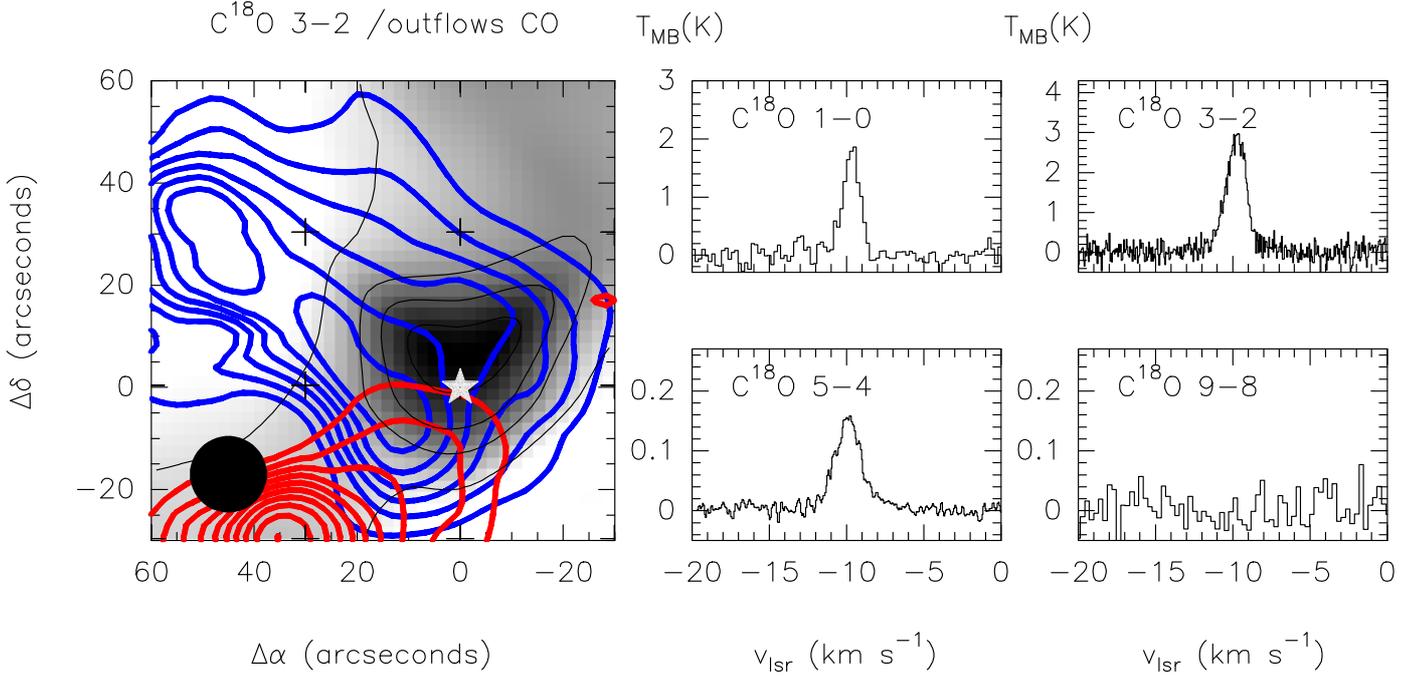}
\vspace{10.0cm}
      \caption{{\it Left}~Integrated intensity map of the C$^{18}$O J=3$\rightarrow$2 line as observed with the JCMT. Crosses mark the observed positions 
and the star indicates the position of FIRS~2
(RA (J2000)=21$^{\rm h}$43$^{\rm m}$01\fs7 Dec(J2000)=66\degr03\arcmin23\arcsec). The grey scale varies linearly 
between 1 K~km~s$^{-1}$ and 5.5~K~km~s$^{-1}$, and the thin black contours 
correspond to 1.5, 3.0, 4.0, and 5.0~K~km~s$^{-1}$. The JCMT beam is shown in the bottom-left corner.
The red and blue contours are adapted from Fig. 1d of Fuente et al. (2001) and trace the high velocity CO J=2$\rightarrow$1 emission,  the contours start 
at 10~K~km~s$^{-1}$ and increase in steps of 5~K~km~s$^{-1}$.{\it Right} 
Spectra of  the C$^{18}$O J=1$\rightarrow$0, J=3$\rightarrow$2, J=5$\rightarrow$4 and J=9$\rightarrow$8 lines towards FIRS~2.}
         \label{Fig 1}
 \end{figure*}

\section{Observations}
The C$^{18}$O 9$\rightarrow$8 line was observed with HIFI on board Herschel on 3 March 2010 (OD 293, obsid 1342191613)
using the wide band spectrometer (WBS) and high resolution spectrometer (HRS)
backends. The observing mode was Double Beam Switch (DBS) fast chop.The HIFI beam size 
corresponds to ∼22$"$ ($\approx$28000~AU at the distance of FIRS~2).
The calibration uncertainty for the HIFI data is of the order
of 20\% and the pointing accuracy is approximately 2$"$ (Roelfsema et al . 2012). The measured
line intensities were converted to main-beam brightness temperatures
using a beam efficiency $\eta_{MB}$=0.74. Data processing started from the standard
HIFI pipeline in the Herschel interactive processing environment
(HIPE)  5.0\footnote{HIPE is a joint development by the Herschel Science Ground
Segment Consortium, consisting of ESA, the NASA Herschel Science
Center, and the HIFI, PACS and SPIRE consortia} (Ott et al. 2011). Further reduction and analysis were
done using the CLASS software (Hily-Blant, P., Pety, J \& Guilloteau, S., 2005). The spectra from the
H- and V-polarizations were averaged in order to obtain a better
S/N. After 21min of integration, the C$^{18}$O 9$\rightarrow$8 line was undetected. The rms noise achieved
in the HRS was 0.034~K with the spectral resolution $\sim$240~kHz. The rms noise in the WBS data is lower than that in the HRS data
by a factor of 1.4 when binned to the same resolution due to the loss factor in
the HRS autocorrelator. 

The JCMT observations of C$^{18}$O 3$\rightarrow$2 were obtained as part of the program M09BC10 in a 15 minute integration on 15 July 2010 using HARP-B, where $^{13}$CO 3$\rightarrow$2 observations 
(not used here) were obtained in the same setting.  The observation was done in stare (point) mode using three repeats, with a rotation of the K-mirror by 90 degrees with each repetition in order 
to fill in gaps caused by missing detectors.  In consequence, the final result of the observation is a grid of 16 spectra situated 30" apart in a 2$'$$\times$2$'$ square region (see the spectra map in Fig. 1). 
Given that the half-power beam width (HPBW) of the JCMT at this frequency is $\approx$ 15", the map is highly undersampled.
Due to the extended nature of FIRS~2, position switch mode was used with a reference position at 
an offset of (720$"$, 720$"$), shared OFF positions were used to increase the efficiency of the observation.
We achieved an rms of 0.195~K with a spectral resolution of $\sim$64~kHz. 

The HDO 3$_{12}$$\rightarrow$2$_{21}$ line at 225.89672~GHz was observed towards FIRS~2 with the IRAM 30m telescope 
(Pico de Veleta, Spain) in 2009, July. The HPBW, forward efficiency and beam efficiency of the telescope at this frequency were 11$"$, 0.91 and 0.63 respectively. As spectrometer we used the 
autocorrelator VESPA configured to provide a spectral resolution of 78 kHz ($\approx$0.10~km~s$^{-1}$ at the line frequency). 
After integrating 137min, the rms noise achieved was 0.02~K. 

\section{Results}
The left panel of Fig.~2 shows the integrated intensity map of the C$^{18}$O 3$\rightarrow$2 line. The emission reveals an elongated structure in the North-South direction, with a peak close to the star position. 
There is a shift of 5$"$ between the position of the C$^{18}$O 3$\rightarrow$2 peak and the position of the protostar as traced by the interferometric continuum 1.3mm observations. Taking into account the spacing 
in the J=3$\rightarrow$2 map (30$"$, approximately two beams), we do not consider this displacement significant.  In the right panel of Fig.~2 we show the spectra of the  C$^{18}$O 3$\rightarrow$2 line towards
the star as observed with the IRAM 30m telescope, the JCMT, and the HIFI spectra of the 5$\rightarrow$4 and 9$\rightarrow$8 lines. In the case of the  C$^{18}$O 1$\rightarrow$0 line, we did not observe towards the star position. The 
spectrum shown in Fig.~2 is derived by averaging the spectra of the closest positions observed in our map, ($-$6.2$"$,$-$8.4$"$), ($-$6.2$"$,$+$11.6$"$), ($+$13.8$"$,$-$8.4$"$), and ($+$13.8$"$,$+$11.6$"$). 
The spectra have been averaged assuming equal weight. This is reasonable taking into account that the emission is quite flat with similar intensity in the four positions 
(see the spectral map in Fig.~3).

\onlfig{3}{
\begin{figure*}
\includegraphics{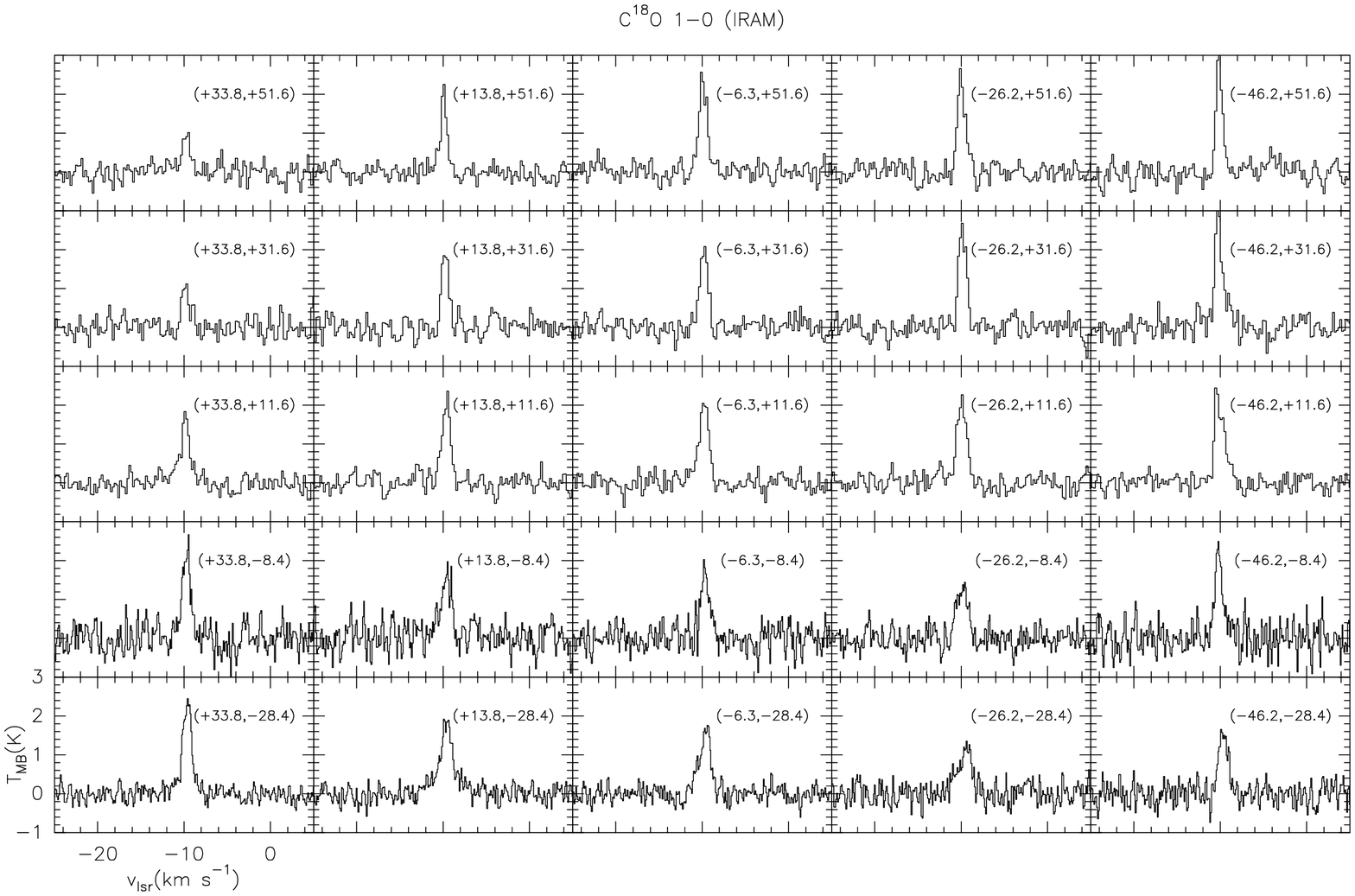}
\vspace{11cm}
      \caption{Spectral map of the C$^{18}$O 1$\rightarrow$0 observed towards FIRS~2 observed with the IRAM 30m telescope. The (0,0) position corresponds 
to 21$^{\rm h}$ 43$^{\rm m}$ 01\fs7 63\degr 03\arcmin 23\arcsec (J2000). These spectra were already presented in Fuente et al. (2001).}
         \label{Fig 1}
 \end{figure*}
}

In Table~1 we show the Gaussian fits to the C$^{18}$O 1$\rightarrow$0, 3$\rightarrow$2 and 5$\rightarrow$4 rotational lines.   The line widths are all found to be narrow ($<$2.0~km~s$^{-1}$)  and are similar to those measured by Fuente et al. (2005b) in well known tracers of cold gas, such as N$_2$H$^+$ 1$\rightarrow$0, CN  1$\rightarrow$0 and H$^{13}$CO$^+$ 1$\rightarrow$0, which 
suggests that the C$^{18}$O emission does not come from the outflow or the hot core. Despite this,
the C$^{18}$O linewidths increase from 1.5~km~s$^{-1}$ in the J=3$\rightarrow$2 line to 2.0~km~s$^{-1}$ in the J=5$\rightarrow$4 line. As discussed in the following Sections,
although some contribution from the outflow cannot be discarded, these linewidths are consistent with the scenario of a collapsing core. 

\begin{table*}
\caption{Summary of observations towards FIRS~2 $^1$}
\label{tab_observations}
\begin{tabular}{llccccc} \hline 
\multicolumn{1}{l}{Line} &  \multicolumn{1}{l}{} & \multicolumn{1}{c}{Area} & \multicolumn{1}{c}{v$_{lsr}$} &
\multicolumn{1}{c}{Width} & \multicolumn{1}{c}{T$_{MB}$} &  \multicolumn{1}{c}{HPBW} \\ 
\multicolumn{1}{l}{} & \multicolumn{1}{c}{} & \multicolumn{1}{c}{(K~km~s$^{-1}$)} & \multicolumn{1}{c}{(km~s$^{-1}$)} &
\multicolumn{1}{c}{(km~s$^{-1}$)} &  \multicolumn{1}{c}{(K)} &  \multicolumn{1}{c}{($"$)} \\ \hline
C$^{18}$O  &  1$\rightarrow$0$^2$ &   2.53(0.10)  &  $-$9.67(0.03)  &   1.29(0.06) &  1.84  & 22 \\
           &  3$\rightarrow$2 &   4.74(0.09)  &  $-$9.76(0.02)  &   1.55(0.04) &  2.86  & 15 \\
           &  5$\rightarrow$4 &   0.31(0.01)  &  $-$9.89(0.02)  &   2.08(0.04) &  0.15 & 42 \\
           &  9$\rightarrow$8 &   \multicolumn{4}{c}{rms = 0.012~K  $\Delta$v=0.33 km~s$^{-1}$}  & 22\\
HDO         &  3$\rightarrow$2 &   0.36(0.04)  &  $-$10.95(0.35) &   6.52(0.87) &  0.05 &  11 \\ \hline
\end{tabular}

\noindent
$^1$ 1-$\sigma$ statistical uncertainties.

\noindent
$^2$ Obtained by averaging the spectra observed towards the positions ($-$6.2$"$,$-$8.4$"$),($-$6.2$"$,$+$11.6$"$),($+$13.8$"$,$-$8.4$"$),($+$13.8$"$,$+$11.6$"$)\\
\end{table*}

In Fig.~4 we show the HDO 3$_{12}$$\rightarrow$2$_{21}$ spectrum observed with the 30m telescope towards FIRS~2.
The Gaussian fit to this line is also shown in Table 1.
Spectrally resolved H$_2$O and H$_2^{18}$O lines towards FIRS~2 were presented and analysed by Johnstone et al. (2010a). 
The H$_2$O profiles show  broad ($\sim$25~km~s$^{-1}$) and  medium ($\sim$6~km~s$^{-1}$) components. In the broad component, the
H$_2$O/CO relative abundance was found to be $\sim$0.2 and the emission was interpreted as coming from the shocked gas produced by the outflows.
The intensities of the medium component ($\sim$6~km~s$^{-1}$) were consistent with the emission arising in the envelope around 
this IM protostar. 
The linewidth of the HDO line is $\sim$6~km~s$^{-1}$, similar to the medium components of the water lines and to the width of the
H$_2$$^{18}$O ground state line. In Fig.~4,
we compare the profile of the 
HDO  3$_{12}$$\rightarrow$2$_{21}$ line (HPBW$\approx$11$"$) with that of the  
p-H$_2$O 1$_{11}$$\rightarrow$0$_{00}$  (HPBW$\approx$19$"$) and the
o-H$_2$$^{18}$O 1$_{10}$$\rightarrow$0$_{01}$(HPBW$\approx$39$"$) lines.

\begin{figure}
\includegraphics{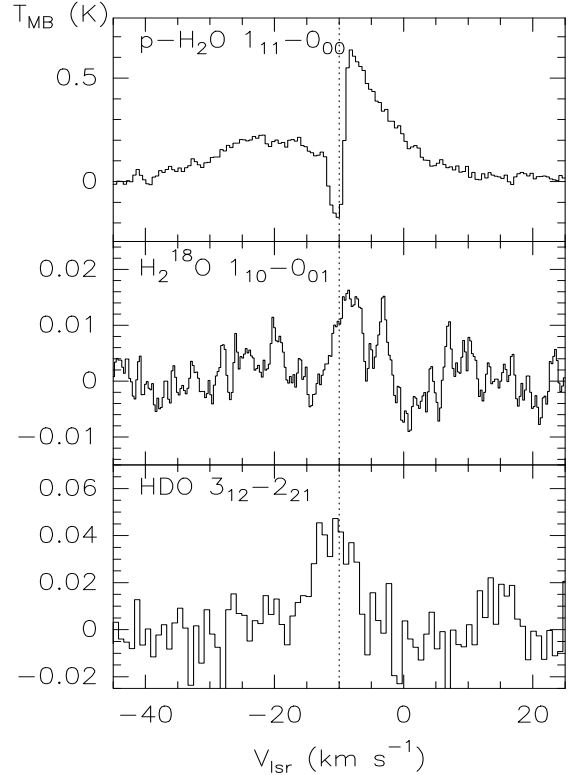}
\vspace{11.0cm}
      \caption{Comparison of the spectra of the p-H$_2$O 1$_{10}$$\rightarrow$0$_{00}$ and o-H$_2$$^{18}$O 1$_{10}$$\rightarrow$0$_{01}$ lines observed with HIFI 
towards FIRS~2 (Johnstone et al. 2010a)  with that of the HDO 3$_{12}$$\rightarrow$2$_{21}$ line observed with the 30m telescope.}
         \label{Fig 2}
 \end{figure}

\section{The C$^{18}$O abundance profile}

The first goal of this paper is to determine the C$^{18}$O abundance profile in the protostellar envelope of FIRS~2. 
Our fitting method consists of two steps.
The C$^{18}$O abundance profile is calculated using the chemical code originally described in 
Caselli et al. (2002), that has been updated as described in Caselli et al. (2008)
to include thermal desorption and also with new measurements of
the CO and N$_2$ binding energies (Collings et al. 2003, {\"O}berg et al. 2005) and sticking coefficients 
(Bisschop et al. 2006). The model inputs include dust temperature and density profiles 
(similar to Emprechtinger et al. 2009). In our chemical model the 
CO and N$_2$ can freeze out and return to the gas phase via thermal desorption and
cosmic-ray impulsive heating (following Hasegawa et al. 1992;
Hasegawa \& Herbst 1993). Other mechamisms to release the molecules from ice mantles such as photo-desorption and/or sputtering in shocks
are not considered. The abundance of the molecular ions
(N$_2$H$^+$, HCO$^+$, H$_3^+$ and their deuterated isotopologues) are calculated
in terms of the instantaneous abundances of neutral species. This 
assumption is based on the short time scale of ion chemistry, compared
to the depletion time scale. This code is well adapted for pre-stellar cores and the envelopes of young Class 0 low mass and IM protostars where the UV radiation
is not expected to play a dominant role on the chemistry.

The density-temperature profile obtained by
Crimier et al. (2010) was used as the physical basis for the chemical model (see Fig.~5). 
Once the C$^{18}$O abundance profile is derived, we use the ray-tracing radiative transfer code DATACUBE to fit 
the C$^{18}$O observations (see Alonso-Albi et al. 2010 for a short description).
Assuming appropiate radial profiles for the temperature, density, molecular abundance, and 
a kinematical structure (defined by the global velocity field and the non-thermal linewidth), this code calculates the brightness temperature distribution on the sky
which is then convolved with the telescope beam to compare with observations. In this way,
beam dilution is naturally taken into account. 
The excitation temperature in each cell is calculated 
using the LVG approximation.
In the first iteration, no systematic global motion was assumed in the envelope and the non-thermal linewidth was adopted 2.5~km~s$^{-1}$. 
As commented later, this has little impact on the derived C$^{18}$O abundances
since the lines are optically thin.

\begin{table*}
\caption{Summary of models}
\label{tab_models}
\begin{tabular}{ll}  \hline
Model & Description \\
\hline 
 1   &   Binding energy for CO = 1100 K \\
 2   &   Binding energy for CO = 5000 K \\
 3   &   Binding energy for CO = 1100 K; only 10\% of the evaporated CO survives at T $>$~100~K \\
 4   &   Binding energy for CO = 1100 K; only 10\% of CO is evaporated back in the gas phase at T$>$~20~K \\
 5   &   Binding energy for CO = 1100 K; t=10$^5$ yr \\
 1*   &   Same as model 1, but with the density increased by a factor of 3  \\
 2*   &   Same as model 2, but with the density increased by a factor of 3  \\
\hline
\end{tabular}
\end{table*}

\begin{figure}
\includegraphics{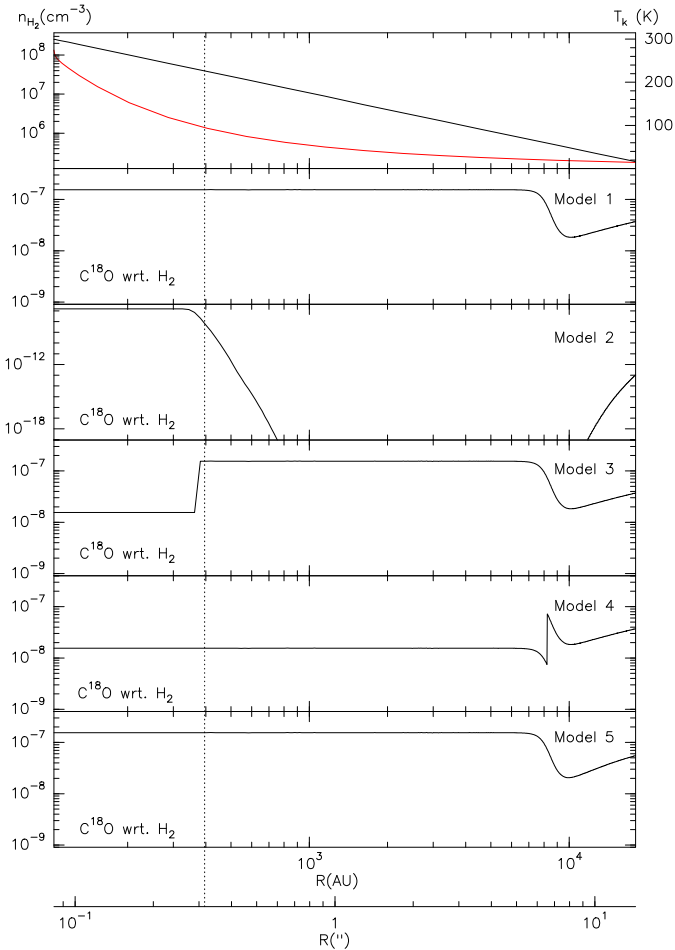}
\vspace{13.0cm}
      \caption{In the upper panel we show the density (black) and temperature (red) profiles (Crimier et al. 2010) used as the physical basis for our chemical model. Below, we show the abundance 
profile of C$^{18}$O predicted by the chemical models in Table 2. The spatial and angular scales are shown. This size of the hot core is indicated by a 
dashed line.}
         \label{Fig 4}
 \end{figure}

The chemical models considered in this Section are described in Table~2 and Fig.~5. 
In Fig.~6 we compare model predictions with the observational results. In the standard model (model 1),
the C$^{18}$O abundance decreases towards the centre of the protostellar envelope, because of  CO depletion onto grain mantles,
until the radius at which gas and dust reach the CO evaporation temperature ($\approx$20-25 K)  where the CO is released back to the gas phase 
and the C$^{18}$O abundance increases sharply to 1.6$\times$10$^{-7}$.
The standard model fails to reproduce the observational results, producing higher intensities in all the observed lines. The greatest discrepancy is found in the C$^{18}$O J=9$\rightarrow$8 line for which model 1 predicts intensities
of $\approx$0.2~K, while we have not detected the line down to an rms of almost a factor of 10 lower. The predicted intensity of the C$^{18}$O J=5$\rightarrow$4 line is also a factor of 5 larger than the observed value and the intensity of the C$^{18}$O J=3$\rightarrow$2 line is larger by a factor of 2. 
Although about 50\% of  the line emission of the J=9$\rightarrow$8 transition is thought to come from the hot core (defined as the region with temperature $>$100~K), 
the contribution of the hot core to the emission of the C$^{18}$O rotational lines with J$<$5 is negligible (see also Y{\i}ld{\i}z et al. 2010). The comparison of the observed lines with model predictions suggests that model 1 overestimates the C$^{18}$O abundance not only in the hot core, 
but in a large fraction of the envelope. Only the intensity of the C$^{18}$O J=1$\rightarrow$0 line  agrees reasonably well with model~1, but this agreement 
could be spurious since the line is contaminated by emission from the foreground molecular cloud. Furthermore, our spectrum is actually an average of the $\sim$20$"$$\times$20$"$ central region.

To account for the observed C$^{18}$O line intensities we need to decrease the C$^{18}$O abundance all across the protostellar envelope. One possibility is that CO is 
locked in a CO-H$_2$O ice matrix with a higher binding energy of 5000~K (model 2). This high binding energy would keep most of the CO frozen onto the grain mantles until 
the dust temperature increases up to $\approx$100~K, i.e., the hot core region. As shown in Fig. 6, model 2 fails to explain all the observed C$^{18}$O lines. The high dust 
temperatures required to release the CO to gas phase ($\approx$100~K) are only found for R$<$400~AU, or 0.3$"$ at the distance of FIRS~2 (see Fig.~5),  and as commented above this small region 
has a negligible contribution to the emission of the J$<$5 rotational transitions.  Model 2 therefore cannot account for all the J$<$5 rotational transitions. We discard therefore
the higher binding energy solution. Note, however, that model 2 still overestimates the intensity of the C$^{18}$O 9$\rightarrow$8 line. This suggests that the C$^{18}$O abundance  
is lower than the reference value even in the hot core.

Model~3 explores the possibility that CO is photodissociated by the central protostar. Like model 1, model 3 assumes the CO binding energy to be 1100~K  but the CO abundance 
is 10\% of the standard when T$_k$$>$100 K (see Fig. 5). Since the region with T$_k$$>$100~K is very small, we obtain essentially the same results as model 1 for the J$<$5 lines, overestimating 
their line intensities. The intensity of the J=9$\rightarrow$8 line decreases, but it is still overestimated since the envelope with T$_k$$<$100~K also emits at this frequency. We need to lower the 
CO abundance also in the T$_k$$<$100~K part of the envelope to fit the observations.

Model~4 assumes that the CO binding energy to be 1100~K  but only 10\% of the initial CO is released back to the gas phase once the ice is evaporated (see Fig. 5). The agreement 
of model 4 with observations is very good for all the lines, both the low-J rotational lines coming from the cold envelope as well as the J=9$\rightarrow$8 line that has a
significant contribution of the hot core. Only the C$^{18}$O 1$\rightarrow$0 line presents some discrepancy, with the observed intensity being twice that predicted. Alonso-Albi et al. (2010) fitted the radial 
distribution of the integrated intensity  of the C$^{18}$O 1$\rightarrow$0, N$_2$H$^+$ 1$\rightarrow$0 and N$_2$D$^+$ 3$\rightarrow$2 lines using the same set of models and they obtained that 
model 4 (model 3 in Alonso-Albi et al.'s paper) gives the best fit to these three maps although the absolute intensity of the C$^{18}$O 1$\rightarrow$0 line towards the star is underestimated. This model 
is therefore our best fit to all the observations.

\subsection{Kinematical structure of the envelope}
Thus far we have used a spherical model with a non-thermal linewidth of $\Delta v$=2.5~km~s$^{-1}$ and without any systematic motion in the envelope. In this section
we consider the more physical scenario of a collapsing envelope. We obtained an excellent fit to the observations assuming the abundances derived in model~4 and that 
the envelope is 
collapsing with a radial velocity law $v_c$=41.8$\times$(R(AU))$^{-0.5}$~km~s$^{-1}$ and a non-thermal linewidth $\Delta v$=1.2~km~s$^{-1}$ that is kept fixed across the 
envelope.
Moreover, our collapse model is able to fit the increase in the linewidth between the J=3$\rightarrow$2 and the J=5$\rightarrow$4 lines without invoking any contribution from 
the molecular outflow (see $``$Model~4-collapse$"$ in Fig. 6).
This kinematical model was previously used by Johnstone et al. (2010a) to fit the narrower component of the water lines
and is consistent with the free-fall velocity profile for a central object of mass 1.1 M$_\odot$.
It is true that an expansion profile with the same velocity law would produce the same fit.
Since the C$^{18}$O rotational lines are optically thin, the infall and outflow scenarios are not easily
discernible. High spatial resolution observations are necessary to unambiguously determine the kinematical structure of the envelope. 

\begin{figure*}
\includegraphics{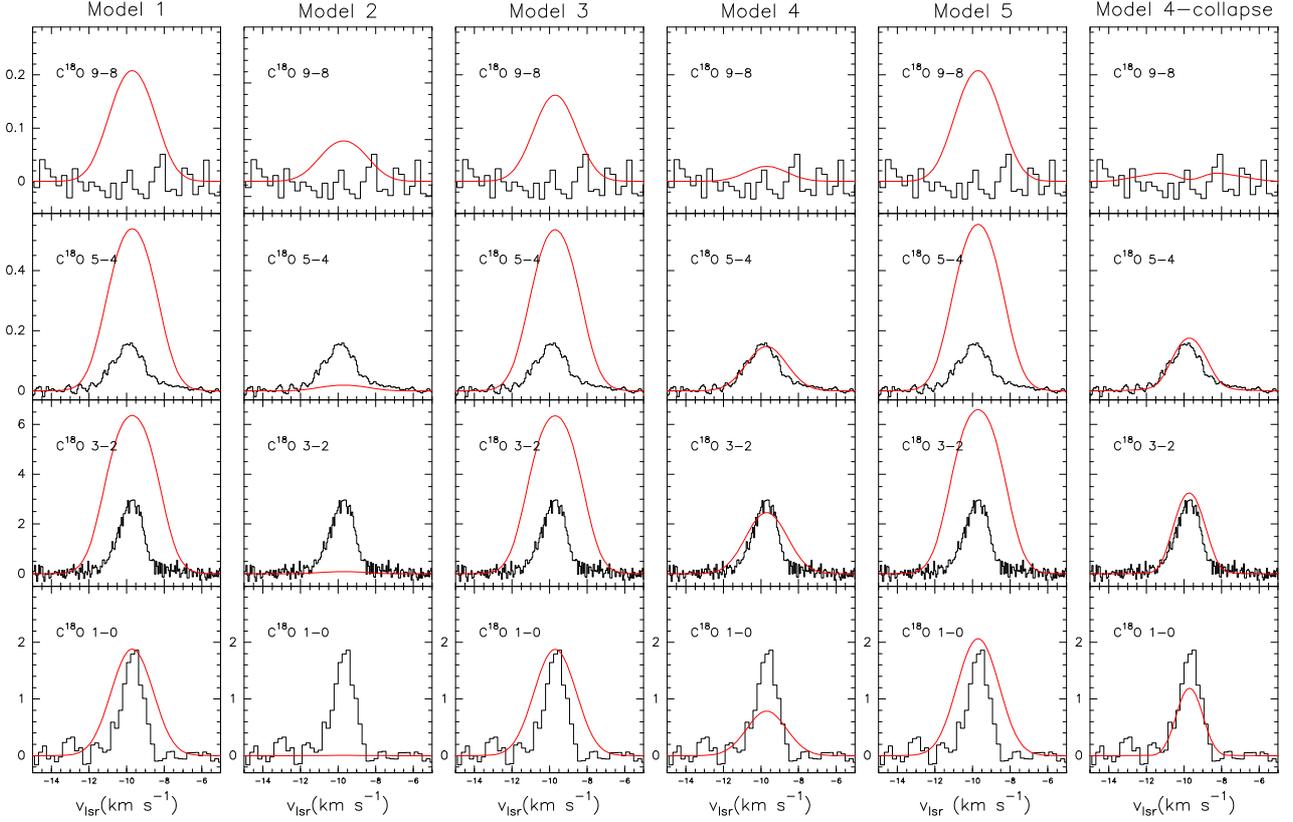}
\vspace{13.0cm}
      \caption{Comparison of models 1, 2, 3, 4, 5 and model 4 assuming a collapsing core with observational results.}
         \label{Fig 3}
 \end{figure*}

\subsection{Uncertainties in our modeling}
Our chemical and radiative transfer models are based on the density-temperature profiles by Crimier et al. (2010). Crimier et al. (2010) adopted the dust opacity calculated by Ossenkopf \& Henning (1994), in particular 
their OH5 dust model, which refers to grains coated by ice. One possibility is that the dust properties in protostellar envelopes are different from those in pre-stellar cores. Grains could have lost their icy mantles because 
of evaporation and shocks.
In this case the opacities at sub-mm wavelengths would be a factor $\sim$3 lower (MRN model in Ossenkopf \& Henning (1994)) than the adopted ones. 
The dust temperature would be little affected by assuming the MRN model since the change in the opacity is quite uniform all over the IR-submm wavelength range. 
Using these opacities, however, the dust density and therefore the dust mass 
obtained from the SED fitting would be a factor of 3 larger.
This situation is investigated in our models 1* and 2* which are the same as models 1 and 2, but with a density a factor of 3 higher 
all across the envelope. The higher dust density produces a higher C$^{18}$O depletion in the cold envelope. However, since the total mass is a factor of 3 larger, the intensities of the C$^{18}$O lines are 
about 3 times larger than in the case of the OH5 opacities, in even larger disagreement with the observed values.

We have assumed that the gas and dust are thermally coupled all over the envelope. Since water is a very efficient coolant, the gas and dust could be thermally decoupled in the inner part of the envelope 
provided that the water abundance is high ($>$10$^{-5}$). Crimier et al. (2010) obtained that for X$_{in}$(H$_2$O)=10$^{-5}$ the difference between the dust and 
gas temperature in the hot core should be $\sim$40~K. To explore the effect of this lower gas temperature could have on the emission of the studied 
C$^{18}$O lines, we have repeated our excitation and radiative transfer calculations assuming that the gas temperature in the hot core is 40~K lower than the dust temperature.
This has a negligible impact in the lines with J$<$5 because the hot core is very small ($\la$0.3$"$). 
The  C$^{18}$O 9$\rightarrow$8 emission arising from the hot core is both thermalized and
optically thin and so for a fixed mass its intensity has a linear dependence on the temperature. In our case, 
the change in temperature is $\la$30~\% (from 152 to 112~K at R=200~AU). Since only half of the flux comes from the hot core, the total
change in the intensity of the C$^{18}$O 9$\rightarrow$8 line is less than 20\%, within the calibration uncertainty. 

One important assumption in our model is that the geometry is simple and the physical structure is unchanged during evolution 
(i.e. we are implicitly assuming that the freeze-out time scale is more rapid than dynamical timescales). Moreover, the freeze-out and desorption 
are followed in time for 1~Myr, whereas the abundances of the molecular ions are calculated on the basis of the instantaneous abundances of CO and 
N$_2$ because of the significantly shorter time scales involved in ion chemistry. The chosen run time is arbitrary (based on rough estimates of 
the time scale of the star formation process as a whole, e.g. Hartmann et al. 2012, so that it could be considered an upper limit for our young 
source). To investigate the impact of this time scale assumption, we have run model 1 and stopped the chemical evolution at t=10$^5$~yr (model 5).
The fit to the observations with the resulting abundances is slightly worse than with model 1. This is easily understood 
taking into account the time scale of the two processes that determine the C$^{18}$O abundance, CO depletion and evaporation. The depletion time at low 
densities (n$\sim$10$^4$~cm$^{-3}$) is $\sim$10$^5$~yr (see e.g. Johnstone et al. 2010b), therefore the CO is not completely depleted in the outer parts of 
the envelope at t=10$^5$~yr. Thermal desorption is a very rapid process and all the CO is back in the gas phase at t=10$^5$~yr. Thus, the global CO abundance is 
larger at t=10$^5$~yr than in the stationary model making the fit slightly worse. 
However, since this increase of the C$^{18}$O abundance occurs in the outer lower density part of the envelope, the change in the emission of the J$>$3 lines is very small and
does not affect our fit. 

One important caveat of our model is that we assume spherical symmetry. It is well known that outflows excavate biconical cavities in the protostellar envelopes. The UV radiation from the star forms 
photodissociation regions (PDRs) in the walls of the cavity. In this case, the PDR would be spatially extended and photodissociation could contribute to lower global C$^{18}$O 
abundance in the protostellar envelope. The relative importance of this mechanism in FIRS~2 is difficult to evaluate without a detailed knowledge of the geometry and density distribution
at small scales. The existence of a hot core rich in complex molecules (Fuente et al. 2005a) and the high deuteration degree measured by Fuente et al. (2005b) suggest that 
the protostar is in a very early evolutionary stage and the chemistry of the bulk of the dense molecular core is not affected by UV radiation. Higher spatial 
resolution observations and more sophisticaded 2-D modeling (Bruderer et al. 2009, 2010; Visser et al. 2012) are required to establish a firm conclusion. 

\section{HDO}

In order to estimate the HDO abundance in this protostellar envelope, we have modeled
the first 25 rotational levels of both ortho- and para-HDO using the same nonlocal code for lines and dust continuum used in the
analysis of H$_2$O toward Sgr B2 by Cernicharo et al. (2006). The code has been described elsewhere (Gonz\'alez-Alfonso \& Cernicharo 1993).
The HDO collisional rates are taken from Green (1989). 
We assume a step function for the HDO abundance profile with X$_{in}$ the HDO abundance for T$_k$$>$100~K and X$_{out}$
the HDO abundance for larger radius. The HDO 3$_{12}$$\rightarrow$2$_{21}$ line has E$_{low}$=153~K, which implies that only 
the hot core gas (T$_k$$>$100~K) significantly contributes to its emission. 
This line provides therefore an excellent opportunity to probe the inner protostellar envelope.

Assuming the infall velocity profile described in Section 4.1 and the density-temperature profile of Crimier et al. (2010), 
the observed HDO 3$_{12}$$\rightarrow$2$_{21}$ line is well reproduced with a X$_{in}$=4$\times$10$^{-8}$ (see Fig.~7, panel A).  
The intensity of the HDO 3$_{12}$$\rightarrow$2$_{21}$ line does not depend on the abundance in the outer envelope.  
We consider two possible scenarios that could limit the abundance of HDO in the outer envelope:  (i) in hot corinos the 
ratio of [HDO]/[H$_2$O] is typically $\approx$0.1, whilst (ii) in massive star forming regions the  [HDO]/[H$_2$O] ratio is more 
often found to be $\approx$0.001.  Johnstone et al. (2010a) estimated a water vapour abundance of 10$^{-7}$ in the outer envelope and this 
gives us X$_{out}$(HDO)=$\sim$10$^{-8}$ for case (i) and X$_{out}$(HDO)=$\sim$10$^{-10}$ for case (ii), respectively.  We obtained in both 
cases an equally good fit using X$_{in}$=4$\times$10$^{-8}$, confirming our expectation that the outer envelope abundance does not significantly
contribute to the HDO line profile.

Another result of our modeling is that under the physical conditions of this IM hot core
(T$_k$$>$100~K, n(H$_2$)$>$5$\times$10$^7$~cm$^{-3}$) the  HDO~3$_{12}$$\rightarrow$2$_{21}$ line is masering
with $\tau$= $-$1.8. This is consistent with previous calculations by Ceccarelli et al. (2010)
which proved that for T$_k$$>$50 the 225~GHz line masers at densities larger than
10$^6$~cm$^{-3}$, with a peak around 10$^8$~cm$^{-3}$ (see Fig.~4 of Ceccarelli et al. 2010). 
Maser emission is more sensitive to the assumed physical conditions 
than in the case of thermal lines. Below we 
estimate the HDO abundance in the hot core and discuss the impact that a small change in the
assumed physical conditions would have in our HDO abundance estimate.

In Fig.~7  we investigate the effect that uncertainty in the hot core physical conditions 
could have on our results. Model B assumes that the gas
temperature remains fixed at 100~K within the hot core. As seen in panel B of Fig.~7,
this temperature decrease in the hot core would decrease the intenisty of the HDO~3$_{12}$$\rightarrow$2$_{21}$ line
by a factor of 2. 
A change in the density has  larger impact on the derived HDO abundance. 
If we assume the Crimier's temperature profile but fix the density of the hot core 
to 5$\times$10$^7$~cm$^{-3}$, the intensity of the HDO line decreases by a factor of $\sim$4 (panel C in Fig. 7)
and we need to increase the X$_{in}$ to $\sim$10$^{-7}$ to fit our observations.
In fact, we do not need to change the density of the whole hot core to have such a dramatic effect.
Just by fixing the density to  5$\times$10$^7$~cm$^{-3}$ for radii between 150~AU and 300~AU, the intensity of the HDO 3$_{12}$$\rightarrow$2$_{21}$ 
would decrease by a factor of $\sim$3 (panel D). This low density is, however, unrealistic.
Fuente et al. (2005a) presented interferometric observations of the continuum emission at 1.3mm and 3mm towards FIRS~2. 
They derived a compact continuum source toward with a size of $\approx$650~AU$\times$900~AU and a total (gas+dust) mass of 2~M$_\odot$. 
This implies an average molecular hydrogen density of $\sim$2$\times$10$^8$~cm$^{-3}$ in this compact core. 
Consequently, a molecular hydrogen density of 5$\times$10$^7$~cm$^{-3}$ is a very conservative lower limit to the density in the hot core.

Masing depends on the coherence length so that
the derived abundance is also sensitive to the velocity structure of the source
Assuming again Crimier's density-temperature profile,
a non-thermal linewidth of $\Delta$v=5~km~s$^{-1}$ in the hot core and the absence of a systematic motion in the envelope,
the predicted intensity of the HDO 3$_{12}$$\rightarrow$2$_{21}$ line is $\sim$0.02 (blue line in panel A) and we need to increase the 
HDO abundance to X$_{in}$=8~10$^{-8}$ to fit our observations.

Therefore, we can conclude that for reasonable physical conditions in the hot core, the HDO abundance in the hot core is between 0.4$-$1~10$^{-7}$.
We would like to comment that the maser emission of the HDO 3$_{12}$$\rightarrow$2$_{21}$ line is also expected to be
sensitive to the geometry of the region. Any departure from the spherical geometry could also produce a different HDO abundance. 

\begin{figure}
\includegraphics{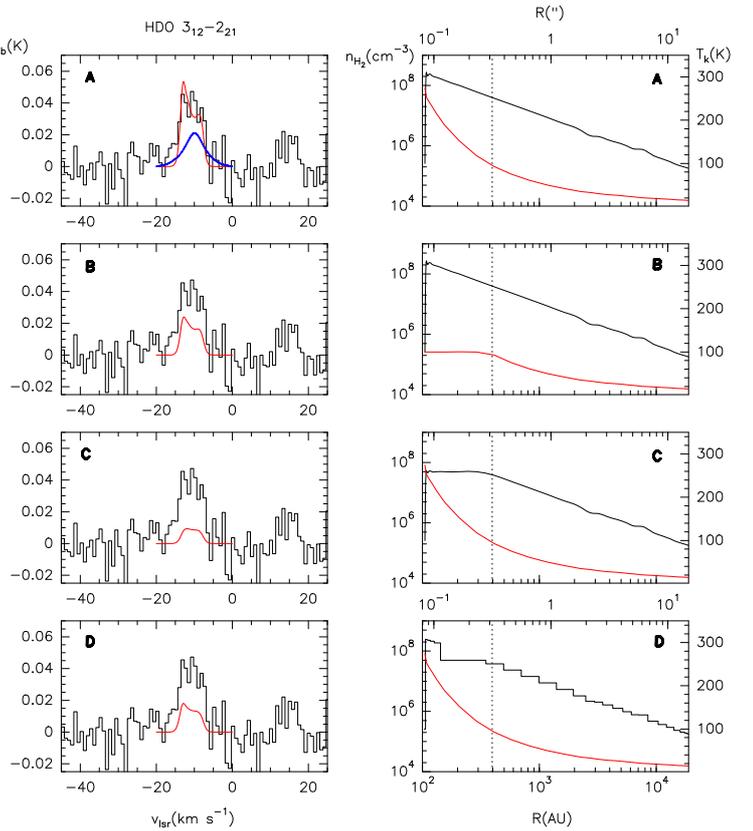}
\vspace{11.0cm}
      \caption{Comparison between HDO 3$_{12}$$\rightarrow$2$_{21}$ line observed with the 30m telescope and the results of our HDO model. On the right we show the density (black) and temperature (red) 
profiles assumed in each case. The angular (upper axis) and spatial (lower axis) scales are indicated. The dashed line marks the extension of the hot core. On the left, we compare the predicted spectra with our observations for each case. {\bf (A)} In this model we adopt X$_{in}$=4$\times$10$^{-8}$ and X$_{out}$=1$\times$10$^{-8}$. 
The red line corresponds to the collapse profile described in Section 4.1. The blue line corresponds to the same abundance profile but with a fixed non-thermal linewidth of $\sim$5~km~s$^{-1}$ and
without any velocity profile. {\bf B} The same as {\bf A} but with the temperature in the hot core fixed to 100~K. {\bf C} The same as {\bf A} but with the molecular hydrogen density in the hot core 
fixed to 5$\times$10$^7$~cm$^{-3}$.{\bf D} The same as {\bf A} but with the molecular hydrogen density in the hot core fixed to 
5$\times$10$^7$~cm$^{-3}$ for radii between 150~AU and 300~AU.}
         \label{Fig 5}
 \end{figure}

\section{Discussion}

There is mounting evidence that the CO abundance is lower than typically expected in protostellar envelopes.
The low CO abundance is not a peculiarity constrained to IM protostars. For example, Y{\i}ld{\i}z et al. (2010) used a combination of ground-based and HIFI data 
to show that the C$^{18}$O abundance must be lower than the standard value by a factor of 3-5 in the evaporation region of the envelope of the low mass protostar in NGC~1333~IRAS~2A. Similarly, Wilson et al. (2001) derived a low CO abundance 
(a factor of 5 below the standard value) in the Orion hot core and in the Orion Ridge on basis of high-J $^{13}$CO lines.  Although the number of protostars in which a physico-chemical modeling of the 
C$^{18}$O emission has been done is still very low the ubiquity of this phenomenon across the stellar mass range, without any obvious dependence on the UV radiation or gas kinetic temperature, is more consistent with its origin being related to the interplay between the gas and grain-surface chemistry.
Alonso-Albi et al. (2010) interpreted this CO deficiency as possible evidence for an active surface chemistry in the dense gas surrounding IM protostars, where CO is efficiently converted into more complex 
organic molecules, such as CH$_3$OH. High spatial resolution interferometric observations revealed a hot core very rich in complex molecules in FIRS~2 (Fuente et al. 2005a). 
The derived abundance of CH$_3$OH, 3$\times$10$^{-8}$, is not high enough to explain the lack of CO, but the emission of this molecule is optically thick and the estimated abundance 
is a lower limit. Another possibility is that the CO is locked in grains in different ice layers with different binding energies resulting in a larger than expected timescale for CO evaporation. This would explain the evolutionary trend that seems to exist in IM protostars, with the lower CO abundance associated with the youngest objects. 

We have derived an HDO abundance of $\sim$0.4$-$1$\times$10$^{-7}$ in the inner part of the envelope of FIRS~2.  This abundance is similar (within a factor of 2) to that found in the hot corinos NGC1333-IRAS 2A (Liu et al. 2011) and IRAS  16293-2422 (Parise et al. 2005) but larger by more than a factor of 100 than the limit obtained in NGC 1333-IRAS 4B (J{\o}rgensen \& van Dishoeck 2010). Fuente et al. (2005a) derived [D$_2$CO]/[H$_2$CO]$\sim$0.14 in the hot core
($\sim$650 $\times$ 900 AU innermost region) towards FIRS~2 based on interferometric observations. They interpreted this high deuteration as the consequence of the evaporation of 
the icy grain mantles. The non-detection of N$_2$D$^+$ towards this hot core reinforced this interpretation (see discussion in Fuente et al. 2005a). In this scenario, the deuterium fractionation of water and formaldehyde could have happened on the grain surfaces during the cold pre-stellar and the collapse phases while the icy mantles are only more recently evaporated. The spread of values of the [HDO]/[H$_2$O] ratio amongst the observed hot cores/corinos and the differences found between the deuterium fractionation of formaldehyde and water
have been recently explained by Cazaux et al. (2011) as the consequence of grain surface chemistry. While the deuterium fractionation of formaldehyde is sensitive to the D/H ratio during the collapse and increases with time and density, the [HDO]/[H$_2$O] ratio depends mostly on
the dust temperature prior to the collapse phase. Following Cazaux et al. (2011) and also assuming an upper limit of $\sim$10$^{-5}$ for the H$_2$O abundance in the hot core,  
the [HDO]/[H$_2$O] ratio found in the hot core of FIRS~2 is consistent with warm dust ($\sim$17~K) prior to collapse.

\section{Conclusions}
FIRS~2 is the most studied and best understood IM protostar and, as such, is an excellent template upon which to base interpretations of  other IM and massive star forming regions.
The wealth of millimetre and interferometric data available for this source and the recent WISH HIFI  observations have allowed us to model the C$^{18}$O  
chemistry and the HDO abundance in this protostar. In particular, the new JCMT and Herschel Space Observatory data prove that the C$^{18}$O abundance in the protostellar envelope and hot core region is lower than 
the standard value, further confirming previous results by Alonso-Albi et al. (2010) based on the low-J rotational lines. This result is not a peculiarity of FIRS~2 chemistry, but seems to be ubiquitous in YSOs in all the mass stellar range. The explanation for this low C$^{18}$O abundance is not yet clear. Grain surface chemistry could play a key role in determining the C$^{18}$O abundance. One possibility is that CO has been efficiently converted to more complex organic molecules in the icy mantles, thus reducing the amount of CO released to the gas phase after evaporation. Energetic shocks and UV radiation illuminating the walls of the cavities excavated by the outflows could also contribute to low the C$^{18}$O abundance. In any case, there seems to exist an evolutionary  trend with the lower CO abundance associated with the youngest objects.

We have modelled the HDO emission to derive an HDO abundance of 0.4$-$1$\times$10$^{-7}$ in the hot core of the IM protostar FIRS~2. This is the first measurement of
the HDO abundance in a IM hot core. The obtained HDO abundance is similar (within a factor of 2) to that found in the hot corinos NGC~1333~IRAS~2A (Liu et al. 2011) and IRAS ~16293-2422 (Parise et al. 2005) but larger by more than a factor of 100 than the limit obtained in NGC~1333~IRAS~4B (J{\o}rgensen \& van Dishoeck 2010) without any systematic trend with the luminosity. The large spread of the [HDO]/[H$_2$O] values has been interpreted by Cazaux et al. (2011) as the consequence of grain surface chemistry.

There is scarce observational data about the abundances of key molecular species such as CO, water and deuterated species in hot core/corinos. 
Moreover, the estimated abundances often suffer from large uncertainties due to the poorly known physical conditions, geometry, even sizes of these objects. To determine the abundance of these species is of paramount importance towards understanding the interplay
between gas phase and grain surface chemistry in protostellar envelopes, and eventually the chemical evolution from hot cores/corinos to protoplanetary disks.


\begin{acknowledgements}
AF and JC have been partially supported within the program CONSOLIDER INGENIO
2010, under grant  CSD2009-00038 "Molecular Astrophysics: The Herschel and ALMA Era –
ASTROMOL". DJ is partially supported by an NSERC Canada Discovery Grant. The WISH team
is thanked for their help and instructive discussions.

\end{acknowledgements}

\end{document}